\documentstyle[preprint,aps]{revtex}

\begin{document}

\title
{Kosterlitz-Thouless Transitions in Bi${}_2$Sr${}_2$CaCu${}_2$O${}_{8+\delta}$ Thin Films by Vortex String}

\author{Tomoko Ota, Ichiro Tsukada, Ichiro Terasaki,\cite{tera} and Kunimitsu Uchinokura}

\address
{Department of Applied Physics, The University of Tokyo, 7-3-1 Hongo, Bunkyo-ku, Tokyo 113, Japan}

\maketitle

\begin{abstract}
Current-voltage ($I-V$) characteristics and the increase of superconducting transition temperature $T_c$ as a function of the number of half-unit-cell layers $n$ are studied on Bi${}_2$Sr${}_2$CaCu${}_2$O${}_{8+\delta}$ thin films grown by molecular beam epitaxy. 
These observations are interpreted in terms of Kosterlitz-Thouless (KT) transition associated with vortex string threading the film. 
Considering the change in the types of the lowest free-energy excitations as a function of the size of vortex pair, we have formulated KT renormalization-group (RG) equations. 
They are with respect to independent vortex and antivortex pairs with distance $r$ less than $n$-dependent characteristic length $r_n$ and then in terms of vortex-string and antivortex-string pairs with $r>r_n$. 
RG equations of these two regions are connected by the continuity of superconducting carrier density and vortex density per area. 
Using experimental data obtained with Bi${}_2$Sr${}_2$CaCu${}_2$O${}_{8+\delta}$ thin films, we have solved the RG equations. 
It is shown that the KT transition occurs in the vortex-string region with interstring $\ln r$ interaction only. 
Our calculation successfully explains KT scaling in $I-V$ characteristics and the relation between $T_c$ and $n$. These results indicate that the interlayer interaction plays an important role in Bi${}_2$Sr${}_2$CaCu${}_2$O${}_{8+\delta}$ thin films.
\end{abstract}


\narrowtext


\section{Introduction}

Two-dimensional (2D) Kosterlitz-Thouless (KT) transition \cite{B,KT,Halperin,H} has been proposed as a possible mechanism for high-$T_c$ superconductivity. Anisotropically Josephson coupled 3D XY model served as a microscopic model for layered superconductors with weak interlayer coupling. \cite{HT,Korsh} In the anisotropic 3D XY model with infinite volume, Hikami and Tsuneto (HT) have proposed two types of excitations, {\it i.e.\/}, independent vortex pairs in each layer and closed vortex rings.\cite{HT} This view is also supported by Korshunov's theory \cite{Korsh} based on the 3D Josephson Junction array. 
Superconducting transition in thin metal films,\cite{Kadin} Bi${}_2$Sr${}_2$CaCu${}_2$O${}_{8+\delta}$ (BSCCO) single crystals,\cite{Martin,Artemenko,Ando,Para} YBa${}_2$Cu${}_3$O${}_{7-\delta}$ (YBCO) single crystal,\cite{Yeh} and Tl${}_2$Ba${}_2$CaCu${}_2$O${}_8$ (TBCCO) single crystal \cite{Kim} were interpreted in terms of 2D KT theory. 2D KT transitions in monolayered or thin-film YBCO have also been observed.\cite{Chen,Teshima,Matsuda92} 
In bulk samples, superconducting transition have been considered to occur owing to KT transitions by independent vortex in each CuO${}_2$ layer despite the presence of interlayer coupling.\cite{Martin,Artemenko,Ando,Para,Yeh,Kim} 

In this paper, we will show by using thin films that excitations of vortex string threading the film do emerge as a result of interlayer interaction. 
Extending HT theory to thin-film cases, we are naturally led to the concept of pairs of vortex string and antivortex string threading the film. 
This concept has been introduced and demonstrated experimentally by Matsuda {\it et~al.} using YBCO thin films prepared by reactive evaporation technique.\cite{Matsuda93} 
They observed the increase in superconducting transition temperature $T_c$ as a function of the number of unit-cell layers and argued that this increase in $T_c$ is due to vortex string and antivortex string. 
However, they did not consider the existence of independent vortex and antivortex pairs in their renormalization-group (RG) analysis. 
This is not always justified.

The purpose of this paper is to show that KT transition by vortex string occurs in BSCCO thin films even when we consider the coexistence of vortex-string and antivortex-string pairs and independent vortex and antivortex pairs. 
It is generally thought that BSCCO is more anisotropic than YBCO. 
Then, the role of independent vortex and antivortex pairs seems more important in BSCCO than in YBCO even if we consider the KT transition by vortex string. 
Consequently, a more rigorous treatment considering the coexistence of both excitations seems necessary. 
Thus, to elucidate the role of interlayer interaction, we study the superconducting properties of BSCCO thin films grown by molecular beam epitaxy (MBE) and obtain the results to be interpreted by KT transition by vortex string. 
We observe the universal jump in the temperature dependence of nonlinear power law $\alpha(T)$ obtained from current-voltage ($I-V$) characteristics and the increase in $T_c$ as a function of the number of half-unit-cell layers $n$ for our BSCCO thin films. 
The increase in $T_c$ as a function of $n$ is rather similar to that of YBCO thin films and can be regarded as the evidence of the presence of vortex strings threading the film. 
 This is in contradiction to the expectation that for very anisotropic BSCCO film the interlayer coupling is so weak that independent vortices may be dominant excitations. 
 We further confirm that the universal jump in $\alpha$-$T$ relation occurs in the vortex-string form and obtain the size of pairs of vortex string and antivortex string. 
This is analyzed on the basis of the KT RG flow with interlayer coupling. 
To consider the strong anisotropy of BSCCO superconductors, we formulate 2D KT theory for multilayered thin film coupled by interlayer Josephson coupling. 
This theory includes three types of low-lying excitations, {\it i.e.\/}, independent vortex and antivortex pairs in each layer, vortex-string and antivortex-string pairs threading the film, and closed vortex rings as proposed by Hikami and Tsuneto.\cite{HT} 
Comparing the free energies of these three types of excitations, we find that the lowest free-energy excitation changes its form from the independent vortex into the vortex string as a function of $r$, the separation between independent vortex (or vortex-string) and antivortex (or antivortex-string) pair. 
We consider the two regions separated by a characteristic length $r_n$. 
$r_n$ will be defined later in Eq.~(\ref{rn}). 
Then, the lowest free-energy excitations are independent vortex and antivortex pairs with distance $r<r_n$, and are vortex-string and antivortex-string pairs threading the film for $r>r_n$. 
Thus, we write KT RG equations with respect to the independent vortex pairs with $\ln r$ intralayer and $r^2$-dependent interlayer interaction for $r<r_n$ and then with respect to vortex-string pairs with $\ln r$ interstring interaction 
for $r<r_n$. 
Differential equations of these two regions are connected by the continuities of superconducting carrier density and the vortex density per area. 
In this way, we have reduced the 3D XY problem to the 2D XY problem with interlayer coupling. 
Solving the 2D KT RG equations in BSCCO thin-film case, we obtain RG flow in which interlayer coupling is considered. 
We observe the transition from independent vortex to vortex string in RG flow at any temperatures. 
We shall see that KT transitions occur in the vortex string region in these RG flows thus obtained. 
Considering the asymptotic behavior of the free energy of a vortex pair we show that the $r^2$ dependent interlayer coupling necessarily leads to the KT transition in the form of vortex string threading the film. 
Analyzing the experimental results on both $I-V$ and $T_c$-$n$ relations consistently, we conclude that the interlayer coupling plays a significant role even in the very anisotropic BSCCO so that the KT transition occurs in the vortex-string excitation. 
Considering the coexistence of the independent vortex, we show that the KT transition by vortex string occurs in films as thick as 20 half-unit-cell layer for the strongly anisotropic superconductors BSCCO. 
Comparing the results of RG flow with $\alpha$-$T$ relation we can determine the size of pairs of vortex string and antivortex string during the KT transition. 
Thus, we believe that our experimental and theoretical method of analysis using thin films will pave the way to a quantitative understanding of the interlayer coupling in superconductors such as BSCCO, YBCO, {\it etc}. 
This will be a step forward to revealing the nature of superconductivity in these materials.

This paper is organized as follows. In Sec. 2, RG analysis of KT transition in thin films is formulated considering interlayer Josephson coupling. 
Using the parameters of BSCCO thin films we will show that KT transition occurs with respect to vortex-string pairs. 
In Sec. 3, the process of sample preparation and the method of measurements are described. 
In Sec. 4, the experimental results on $I-V$ characteristics and the $T_c$ vs $n$ relation  are described and interpreted in terms of KT theory in which vortex string plays the role of vortex in pure 2D KT theory. 
In Sec. 5, the discussion and the conclusion are given.

\section{Renormalization Group Analysis of KT Transition in Thin Films}

KT transition in 2D superconductors is related with binding and unbinding of quantized vortex and antivortex due to the logarithmic interaction between them. 
This transition occurs at a temperature lower than Ginzburg-Landau (GL) mean-field transition temperature $T_{c0}$, below which condensate wave function is formed. In high-$T_c$ superconductors, the effect of interlayer coupling should be considered. 

Now we consider three types of low-energy excitations in 2D XY model with Josephson interplane coupling. 
On the basis of anisotropic 3D XY model with infinite volume, the effect of interplane coupling is considered by Hikami and Tsuneto.\cite{HT} 
According to them, the energy of an independent vortex and antivortex pair in each layer is given by \cite{HT}
\begin{equation}
U_{vp}(r)=2\pi K_1 \ln (r/{\xi}_{ab}(T)) +2\pi K_{\perp}(r/d)^2 +2E_c. 
\label{Uvp}  
\end{equation} 
Here, $K_1$ is the coupling constant in the $ab$ plane, $K_{\perp}$ is the coupling constant along the $c$ axis, ${\xi}_{ab}(T)$ is GL coherence length in the $ab$ plane at temperature $T$  and is defined as ${\xi}_{ab}(T)={\xi}_{ab}(0)(1-T/T_{c0})^{-1/2}$, $d$ is the interlayer distance, and $E_c$ is the core energy to create a vortex in one layer. 
The second term in Eq.~(\ref{Uvp}) expresses the interlayer Josephson coupling energy. 
In the presence of interlayer coupling, Hikami and Tsuneto showed that there are closed vortex rings within a sample. The energy of a vortex ring in an infinite system is given by \cite{HT}
\begin{equation}
U_{ring}(r)=(r/d)(2\pi K_1 \ln (r/{\xi}_{ab}(T)) +2E_c). 
\label{Ur}  
\end{equation} 
$r/d$ is, roughly speaking, the number of the layers which the vortex ring crosses. 
$U_{vp}(r)=U_{ring}(r)$ yields a characteristic length defined as $r_0 =d(K_1/K_{\perp})\ln (r_0/{\xi}_{ab}(T))$.\cite{HT} 
Here, we ignore $E_c$ since it is small. 
The anisotropy ratio $K_1/K_{\perp}$ is defined as $K_1/K_{\perp}=({\xi}_{ab}(T)/{\xi}_c(T))^2$,\cite{HT} where ${\xi}_c(T)={\xi}_c(0)(1-T/T_{c0})^{-1/2}$. Then, (a) for $r<r_0$, independent vortex pairs in each layer dominate. 
(b) For $r>r_0$, closed vortex rings in a 3D volume dominate.

Application of HT theory to thin film leads to the concept of vortex string as follows. 
For BSCCO, we have $r_0$($T$=64.5 K)=1 $\times 10^4$ nm, using the physical parameters in Table \ref{table1}. 
Then for our $n$-layered films ($n\leq 20$), the relation $r_0>nd$ is satisfied. In this case, closed vortex rings cannot exist within the films. 
Then, there can be only pairs of vortex and antivortex strings threading the film and those of independent vortex and antivortex. 
The interaction energy of a vortex-string and antivortex-string pair in $n$-layered film is given by \cite{Matsuda93}
\begin{equation}
U_{st}(r)=n(2\pi K_1 \ln (r/{\xi}_{ab}(T)) +2E_c),
\label{Ust}  
\end{equation}
which is free from $r^2$-dependent interlayer coupling term and less than $U_{ring}(r)$ for $r>nd$. 
Considering the entropy change for multilayered system, 
we have the free energy of the independent vortex pair as
\begin{eqnarray}
F_{vp}(r)=2\pi K_1 \ln (r/{\xi}_{ab}(T))+2\pi K_{\perp}(r/d)^2 \nonumber \\
+2E_c-k_BT\ln(n). \label{Fvp}  
\end{eqnarray} 
The fourth term is the entropy term for multilayered system which is to be understood as the difference from that of vortex string. 
The origin of the entropy term of Eq.~(\ref{Fvp}) will be explained later. 
Equating $F_{vp}(r)$ with $U_{st}(r)$, we introduce a characteristic length $r_n$ defined by 
\begin{equation}
r_n=d \sqrt{
{{(n-1)K_1}\over {K_{\perp}}} \ln \left({{r_n}\over {{\xi}_{ab}(T)}}\right)+{{k_BT \ln (n)}\over{K_{\perp}}} }. \label{rn}
\end{equation}
This is to be distinguished from $r_n$ defined from the relation $U_{vp}(r_n)=U_{st}(r_n)$ by Matsuda {\it et.al.}.\cite{Matsuda93} 
In Eq.~(\ref{rn}), $E_c$ is ignored again. 
For $r<r_n$ the favored excitations are independent vortex and antivortex pairs, and for $r>r_n$ they are vortex-string and antivortex-string pairs. 
We also introduce another parameter 
$l_n = \ln(r_n/{\xi}_{ab}(T))$. 
The values of $l_n$ ($r_n$) are listed in Table \ref{table1}. 
These values of $l_n$ are obtained by using renormalized $K_1$ as will be described later.

The transition from independent vortex pairs to vortex-string pairs can be formulated in 2D KT theory with interlayer coupling. 
We find the lowest free-energy excitation among the three is independent vortex pair for $r<r_n$ and vortex-string pair for $r>r_n$ under the following conditions, 
\begin{eqnarray}
K_1/K_{\perp}&\gg & K_p,~(nd<{\xi}_{ab}(T)) \label{nd1} \\
K_1/K_{\perp}&\gg &K_r,~(nd>{\xi}_{ab}(T)) \label{nd2}
\end{eqnarray}
where
\begin{eqnarray}
K_p&=&(5.436/(n-1))({\xi}_{ab}(T)/d)^2, \label{kp} \\
K_r&=&n^2/((n-1)\ln (nd/ {\xi}_{ab}(T))). \label{kr} 
\end{eqnarray}
For BSCCO film with $n$=20, we have $K_p$=1.14, $K_r$=9.14 and $K_1/K_{\perp}$=10${}^3$ using the parameters in Table \ref{table1}. 
Then, the inequality $K_1/K_{\perp}\gg K_p$ and $K_r$ is satisfied. 
On this basis, we formulate 2D KT theory with respect to independent vortex pair with energy $U_{vp}(r)$ of Eq.~(\ref{Uvp}) for $r<r_n$ and that with respect to vortex-string pair with energy $U_{st}(r)$ of Eq.~(\ref{Ust}) for $r>r_n$. 
Then, the differential equations of the two regions are connected. 
We will find that the KT transition occurs in the vortex-string region with interstring $\ln r$ interaction of $U_{st}(r)$. 
After solving the RG equations, we will reconfirm in Sec.~5 that the lowest free-energy excitations are such as we have supposed initially. 
A general proof for the absence of dissociation of independent vortex and antivortex pairs will be also given in Sec.~5.

Now, we introduce in advance the various transition temperatures to be used in this paper. 
$T^{\ast}_{KT}$ is the KT-transition temperature of a single half-unit-cell layer system taken as a scaling parameter for $T$. 
$T_{c}(n)$ is the superconducting transition temperature for $n$-layered system at which the measured resistance vanishes. 
$T'_{c}(n)$ is the superconducting transition temperature for $n$-layered system calculated from the RG equation. 
$T_{c0}(n)$ is the Ginzburg-Landau (GL) mean-field transition temperature for $n$-layered system. 
$T_{mid}(n)$ is the midpoint temperature for $n$-layered system. 
$T_{mid}(n)$ is regarded as $T_{c0}(n)$. 
These values are listed in Table \ref{table1}.

The starting equations for our KT RG scheme are written as 
\begin{eqnarray}
\epsilon (r+dr)&=&\epsilon (r) +4\pi {{r^2}\over{2k_BT}}2\pi rdrP(r),
\label{rg1} \\
P(r+dr)&=&P(r) 
\exp \left ( -{{dU(r)}\over{k_BT}}\right),
\label{rg2}
\end{eqnarray}
where $dU(r)$ is given by 
\begin{eqnarray}
dU(r)&=&dU_{vp}(r)\equiv
\left ( {{2}\over{\epsilon (r)r}}+{{4\pi K_{\perp}r}\over {d^2}}\right  )dr,
\nonumber \\ 
&& {\rm (independent~vortex)}
\label{dUvp}\\
dU(r)&=&dU_{st}(r)\equiv 
\left ( {{2}\over{\epsilon (r)r}}\right  )dr, \nonumber \\
&& {\rm (vortex~string)} 
\label{dUst}
\end{eqnarray}
and $k_B$ is the Boltzmann constant. 
We explain the notations in Eqs.~(\ref{rg1})-(\ref{dUst}) in terms of Coulomb-gas model. 
$\epsilon (r)$ is the dielectric constant of the medium between unit charges separated by the distance $r$, and $P(r)$ is associated with the density per area of thermally excited pairs of positive and negative charges (vortex and antivortex pairs). 
In Eq.~(\ref{rg1}), the increments of $\epsilon (r)$ is given by the number of dipoles in the region surrounded by circles with radius $r$ and $r+dr$. 
The term ${2}/(\epsilon (r)r)$ in Eqs.~(\ref{dUvp}) and (\ref{dUst}) is the dipolar interaction term and the second term in Eq.~(\ref{dUvp}) is the interlayer Josephson coupling. 
We have considered the screening only on dipolar interactions but not on the interlayer Josephson coupling. 
Dielectric constant $\epsilon (r)$ is defined from renormalized coupling constant in the $ab$ plane $K(r)$ as $\pi \epsilon (r)=1/K(r)$. 
The coupling constant $K(r)$ is defined by 
\begin{eqnarray}
K(r)&=&K_1(r),~{\rm (independent~vortex)}\label{kk1} \\
K(r)&=&nK_1(r).~{\rm (vortex~string)}\label{knk1} 
\end{eqnarray}
The renormalized coupling constant for a single layer $K_1(r,T)$ is defined as 
\begin{equation}
K_1(r,T)={{{\hbar}^2 n_s^{2D}} \over {4m^{\ast} \chi(r,T)}}.\label{k1ns2d}
\end{equation}
This equation is deduced from intervortex interaction energy as calculated by GL formalism in 2D. 
$n_s^{2D}$ is the superconducting carrier density per area in a single 2D plane defined in terms of 3D carrier density $n_s^{3D}$ as $n_s^{2D}=n_s^{3D}d$, $m^{\ast}$ is the effective mass of superconducting electron. 
The factor $n$ in Eq.~(\ref{knk1}) arises since the carrier density associated with vortex string is $n$ times that of independent vortex in a single layer. 
$\chi(r,T) $ is the relative dielectric permittivity of medium in the 2D Coulomb-gas model in a single layer. 
For $r={\xi}_{ab}(T)$, we have $\chi(r={\xi}_{ab}(T),T)$=1 and then $\chi(r,T)$ represents the effect of screening by the renormalization described by Eqs.~(\ref{rg1}) and (\ref{rg2}). 

At $r=r_n$, the differential forms of Eqs.~(\ref{rg1}) and (\ref{rg2}) are replaced by the continuity equations for $\epsilon (r,T)$ and $P(r,T)$. At first, we assume that 
$\chi(r,T)$ is continuous at $r=r_n$ as 
\begin{equation}
\chi(r_n+0,T)=\chi(r_n-0,T),\label{chic}
\end{equation}
and also assume the conservation law for superconducting carrier density per area per layer
\begin{equation}
n_s^{2D}(r_n+0,T)=n_s^{2D}(r_n-0,T)={\rm const}.\label{nsc}
\end{equation}
Here $\pm 0$ denotes the infinitesimal increment or decrement to $r_n$. 
Combining Eq.~(\ref{nsc}) with Eqs.~(\ref{kk1})-(\ref{k1ns2d}) yields the continuity of $K_1(r,T)$ as $K_1(r_n+0,T)=K_1(r_n-0,T)$ and the connectivity of $K(r,T)$ as $K(r_n+0,T)=nK(r_n-0,T)$. 
Thus, we have the connecting equation of $\epsilon (r)$ as
\begin{equation}
\epsilon (r_n+0,T)=\epsilon (r_n-0,T)/n.\label{epsc}
\end{equation}

The connectivity of $P(r)$ at $r=r_n$ is obtained as follows. 
A single independent vortex in a single layer is transmuted into a vortex string threading the film. 
Therefore, in $n$-layered film, the density of the vortex string per area in a film must be $n$ times the density of independent vortex per area in a single layer. 
Thus, we have 
\begin{equation}
P(r_n+0,T)=nP(r_n-0,T).\label{pc}
\end{equation}
Putting the relation
$P(r_n+0,T)\propto\exp( -U_{st}(r_n+0,T)/(k_BT))$ and 
$P(r_n-0,T)\propto\exp( -U_{vp}(r_n-0,T)/(k_BT))$ into Eq.~(\ref{pc}), we have 
\begin{equation}
U_{st}(r_n+0,T)=U_{vp}(r_n-0,T)-k_BT\ln (n). \label{rrnn}
\end{equation}
The right-hand side of Eq.~(\ref{rrnn}) is $F_{vp}(r)$ of Eq.~(\ref{Fvp}). 
The $\ln n$ term of Eq.~(\ref{rrnn}) arises from the entropy difference which in turn arises from the transition from independent vortex to vortex string. 
These results suggest that we must compare the free energy of each excitation in selecting the favored excitations of $n$-layered system.

Now, to solve the RG Eqs.~(\ref{rg1}) and (\ref{rg2}), we introduce the variables \cite{KT,Halperin}
\begin{eqnarray}
l = \ln \left({r\over {{\xi}_{ab}(T)}}\right), \label{l} \\
x(l,T) = {{2k_BT}\over {\pi K(l,T)}} -1, \label{x} \\
y^2(l,T) = r^4 P(r). \label{y}
\end{eqnarray}
$K(l,T)$ is used in place of $K(r,T)$. 
By introducing the variables $l$, $x(l,T)$, and $y(l,T)$, we can rewrite Eqs.~(4) and (5) in terms of Coulomb-gas model in a picture of vortex. Thus, the RG equations~(4) and (5) are transformed into the form, 
\begin{eqnarray}
{{dx(l,T)}\over {dl}}&=& 8{\pi }^2y^2(l,T), \label{xy1}  \\
{{dy(l,T)}\over {dl}}&=& {{2y(l,T)} \over {x(l,T)+1}}\times \nonumber
\end{eqnarray}
\begin{equation}
\left(x(l,T)-{{2K_{\perp}(l,T){\xi}^2_{ab}(T) \exp{(2l)}}\over {K(l,T)d^2}}\right). \label{xy2}
\end{equation}
2D KT RG equations of Eqs.~(\ref{xy1}) and (\ref{xy2}) are written with respect to independent vortex and antivortex pairs in a single layer for $l<l_n$ and with respect to vortex-string and antivortex-string pairs threading the film for $l>l_n$. 
These equations are written in the same form except the coupling constant $K(l,T)$ and $K_{\perp}(l,T)$ defined as follows, 
\begin{eqnarray}
K(l,T)&=& \left\{ \begin{array}{ll}
                        K_1(l,T),  &(l<l_n) \\ 
                        nK_1(l,T), &(l>l_n) \\
                  \end{array} \right.  \label{kpal} \\
K_{\perp}(l,T)& & \left\{ \begin{array}{ll}
                        \neq 0,  &(l<l_n) \\ 
                        =0.      &(l>l_n) \\
                          \end{array} \right. \label{kper}
\end{eqnarray}
The second term in the right-hand side of Eq.~(\ref{xy2}) for $l<l_n$ is an additional term to the usual 2D KT RG equations and represents the interlayer coupling. 
Introducing the variable for a single layer 
$x_1(l,T) = {{2k_BT}\over {\pi K_1(l,T)}} -1$, we have the relation 
\begin{eqnarray}
x(l,T) &=& x_1(l,T),~(l<l_n)\label{xi} \\
x(l,T)+1 &=& (x_1(l,T)+1)/n.~(l>l_n)\label{xs}
\end{eqnarray}

The initial conditions for $x(0,T)$ and $y(0,T)$ in the RG differential equations (\ref{xy1}) and (\ref{xy2}) are derived from the fact that initially at $l$=0 only independent vortex and antivortex pairs in each layer exist for the case $K_{\perp}/K_1\ll$1. 
The initial condition will be reconsidered in detail in Sec.~5 taking account of the entropy corrections in the system. 
Then, $x(0,T)$ is given by \cite{KT,Halperin}
\begin{equation}
x(0,T) = x_1(0,T)= {{2k_BT}\over {\pi K_1(0,T)}} -1. \label{x01}
\end{equation}
Here, $K_1(0,T)$ is written in terms of magnetic penetration length ${\lambda}_{ab}(T)$ as 
$K_1(0,T)=({\hbar}^2dc^2{\epsilon}_0)/(4e^2{\lambda}^2_{ab}(T))$.\cite{Kadin,Martin} 
${\epsilon}_0$ is the dielectric constant of vacuum, $e$ is the unit charge, and $c$ is the velocity of light. We assume ${\lambda}_{ab}(T)={\lambda}_{ab}(0)(1-T/T_{c0}(n))^{-{1\over 2}}$, and ${\lambda}_{ab}(0)$=260 nm.\cite{Shiba} 
Now, we rewrite Eq.~(\ref{x01}) as 
\begin{equation}
x(0,T)
={{8k_Be^2{\lambda}^2_{ab}(0)}\over{\pi {\hbar}^2dc^2{\epsilon}_0}}
{{T}\over{(1-T/T_{c0}(n))}}-1.
\label{x02}
\end{equation}
The initial value for $y(0,T)$ is given in terms of the initial value $x(0,T)$ and the parameter $B$ by
\begin{eqnarray}
2x(0,T)-2\ln {(x(0,T)+1)}-4{\pi}^2y^2(0,T) \nonumber \\
=- B (T/T_{KT}^{\ast}-1).
\label{y01}
\end{eqnarray}
Following 2D KT theory, we assume the linear $T$ dependence in the right-hand side of Eq.~(\ref{y01}). 
Here, the parameter $B$ is defined as \cite{KT,Halperin,H}
\begin{equation}
B \sim {1\over b}{{T_{KT}^{\ast}}\over {T_{c0}(n)-T_{KT}^{\ast}}},
\label{B}
\end{equation}
In our model, the parameters $b$ and $T_{c0}(n)$ are determined from the experimental data as listed in Table \ref{table1}. 
We have used single layer $T_{KT}^{\ast}$ for KT-transition temperature in Eq.~(\ref{B}). 
$T_{KT}^{\ast}$ is taken as a scaling parameter for $T$ to determine the initial values $x(0,T)$ and $y(0,T)$ for a given temperature $T$. 

The connection between the differential equation of the two regions, {\it i.e.\/}, the region $l<l_n$ and the region $l>l_n$, is expressed as continuities of $K_1(l,T)$ and $x_1(l,T)$ at $l=l_n$. 
Then, the connectivity condition in terms of $x(l,T)$ is written as 
\begin{equation}
x(l_n+0,T) +1=(x(l_n-0,T) +1)/n. \label{xc}
\end{equation}
The connectivity condition for $y(l,T)$ is written as 
\begin{equation}
y(l_n+0,T)=\sqrt{n}y(l_n-0,T). \label{yc}
\end{equation}
Equation~(\ref{yc}) is obtained by inserting the connectivity of $P(r,T)$ of Eq.~(\ref{pc}) into Eq.~(\ref{y}). 
At the KT-transition temperature, and for the limit $l\rightarrow \infty $, we have $x(l,T'_c(n))\rightarrow 0$ and $y(l,T'_c(n))\rightarrow 0$. 
Then, from Eq. (\ref{xs}), we have $x_1(l,T'_c(n))\rightarrow n-1$.

Using the parameters of BSCCO films listed in Table \ref{table1}, the RG differential equations of (\ref{xy1}) and (\ref{xy2}) for $x(l,T)$ and $y(l,T)$ were solved with respect to the variable $l$ for the case $n$=1, 3, 6, and 20. 
The parameters $l_n(T'_c(n))$ are derived, using the parameter $b$, $T_{c0}(n)$, and renormalized $K_1/K_{\perp}$ at $T'_c(n)$. 
The parameters $T_{c0}(n)$, and $b$ are determined from the experiment as will be discussed in Sec. 4 (Fig.~\ref{fig2}). 
The initial value of the anisotropy parameter $K_1/K_{\perp}$ is obtained as $K_1/K_{\perp}=10^3$ using the parameters used in Table \ref{table1}. 
However, it is reduced from the initial value by renormalization as $l$ increases. 
Renormalized $K_1(l,T)$ is used in each step of discretized $l$ during the numerical calculation. 
We will confirm in Sec.~5 that the lowest free-energy excitations are such as we have assumed initially even after the renormalization of $K_1(l,T)$.

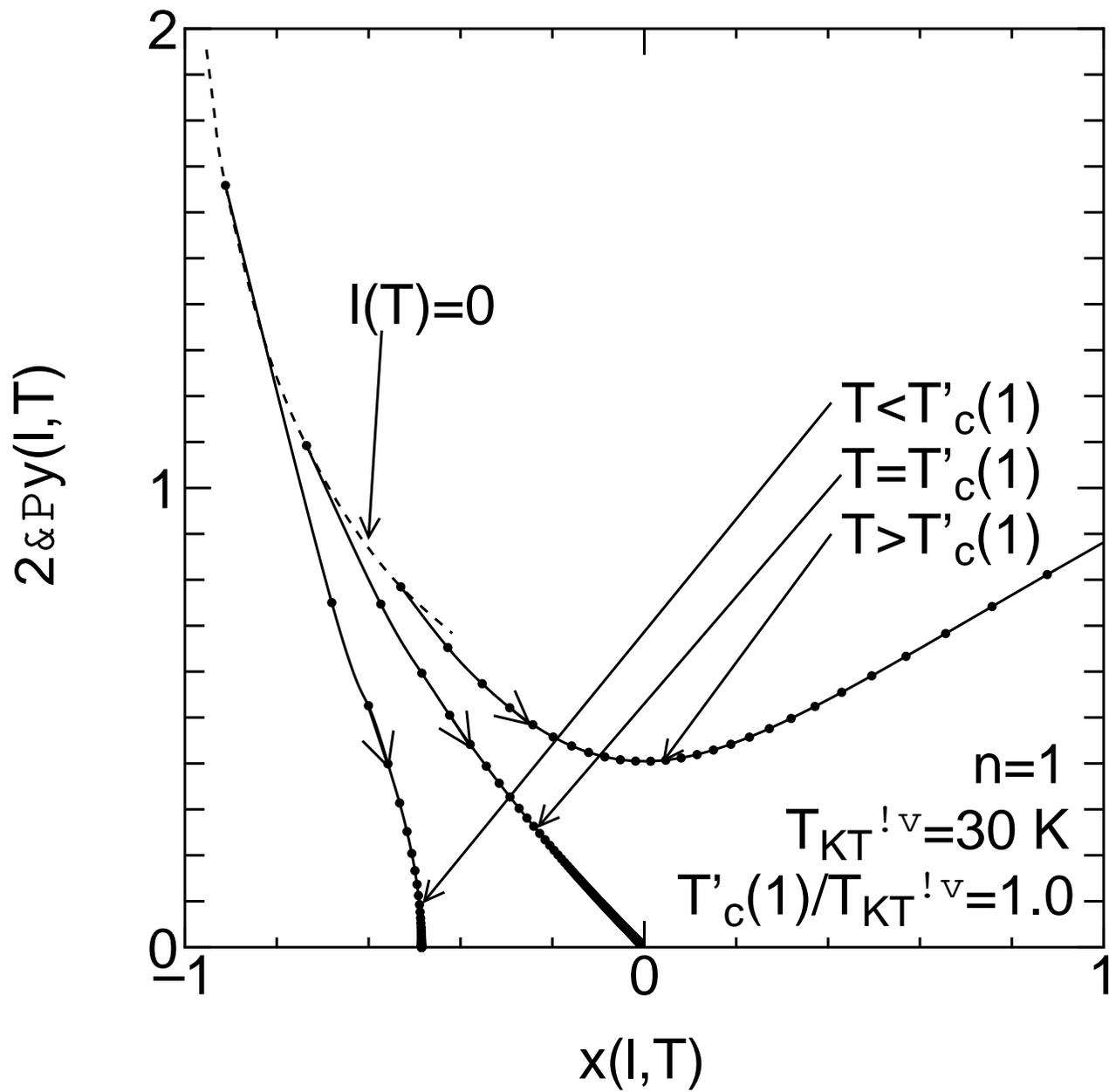
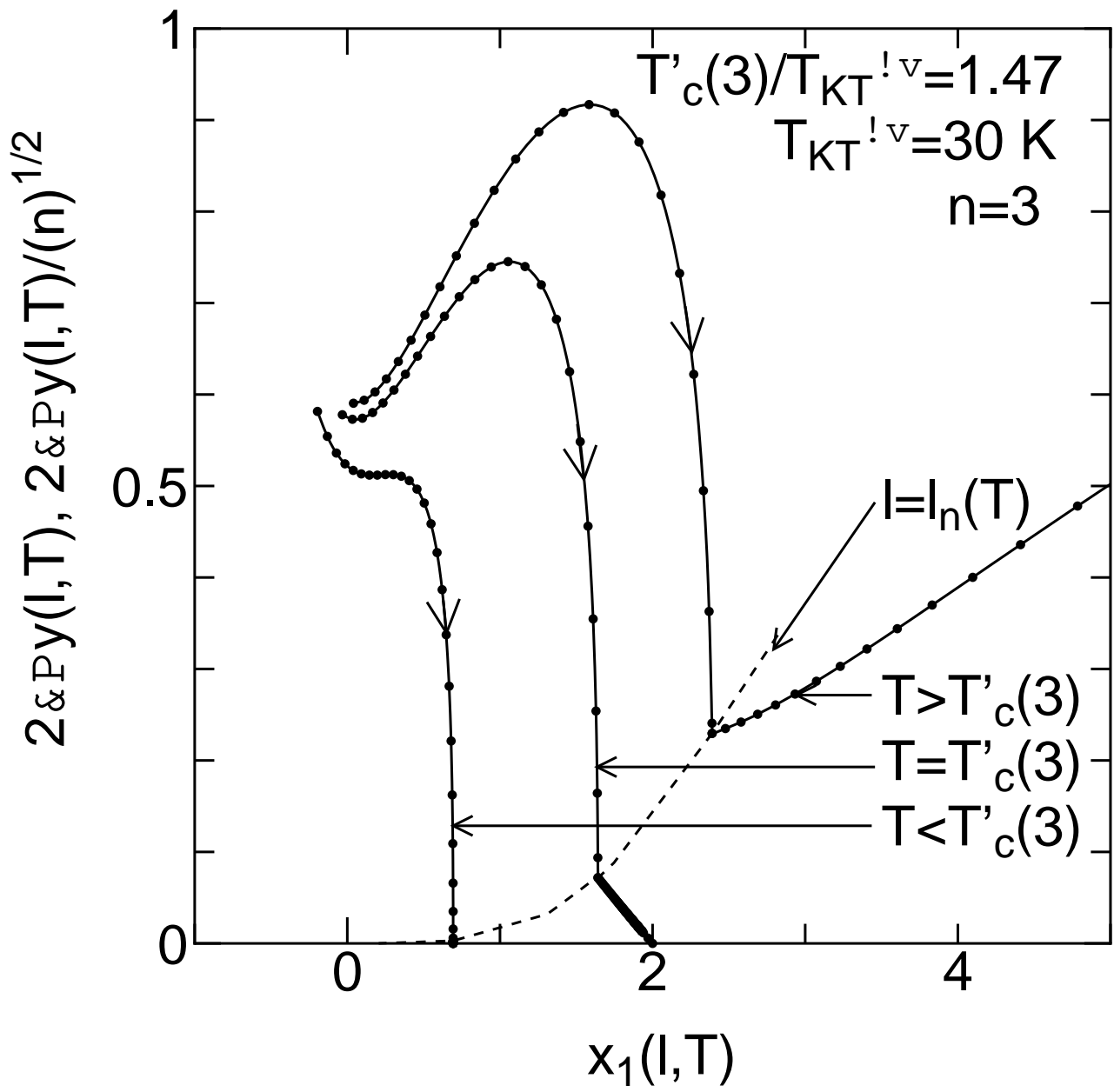
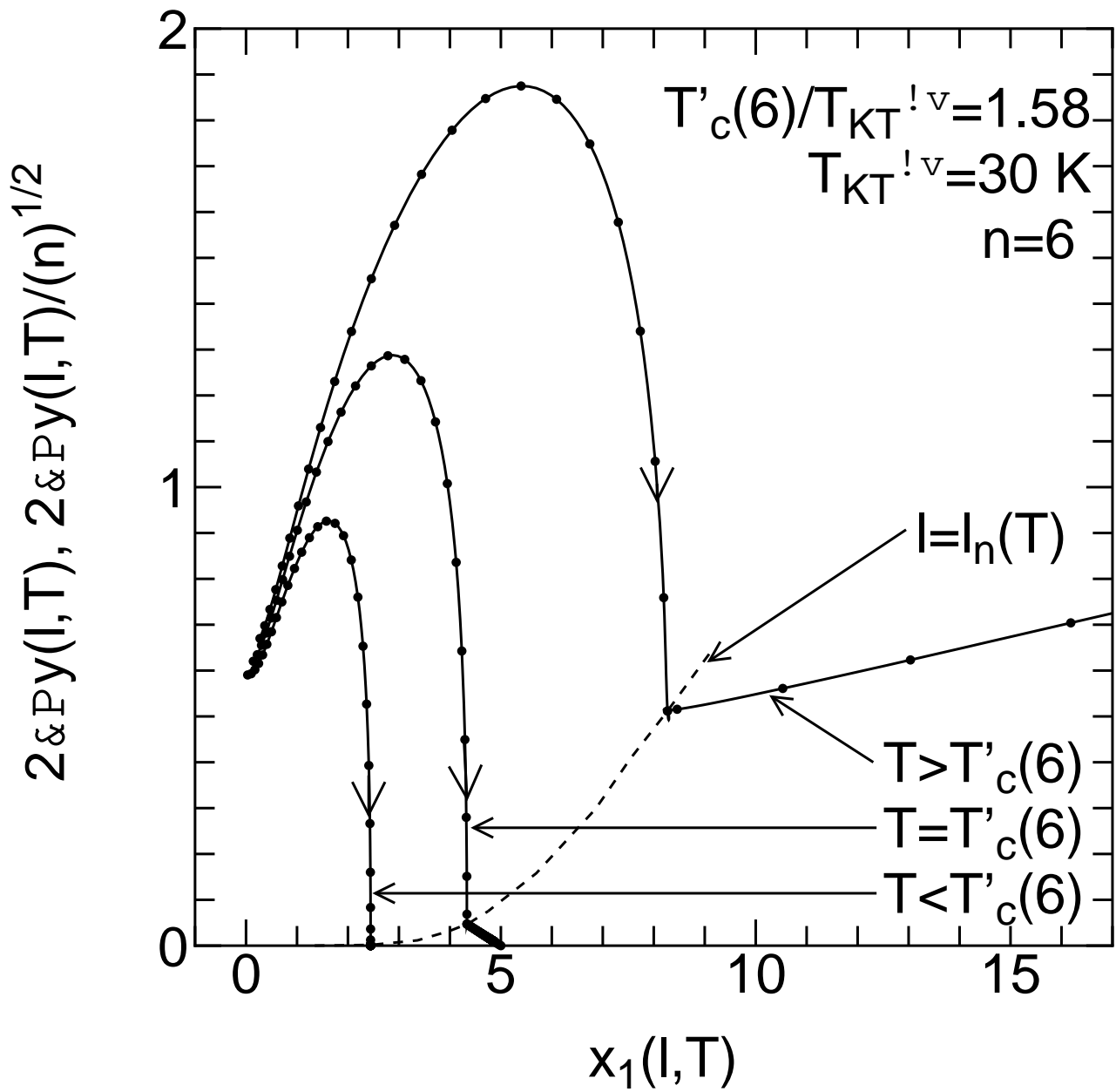
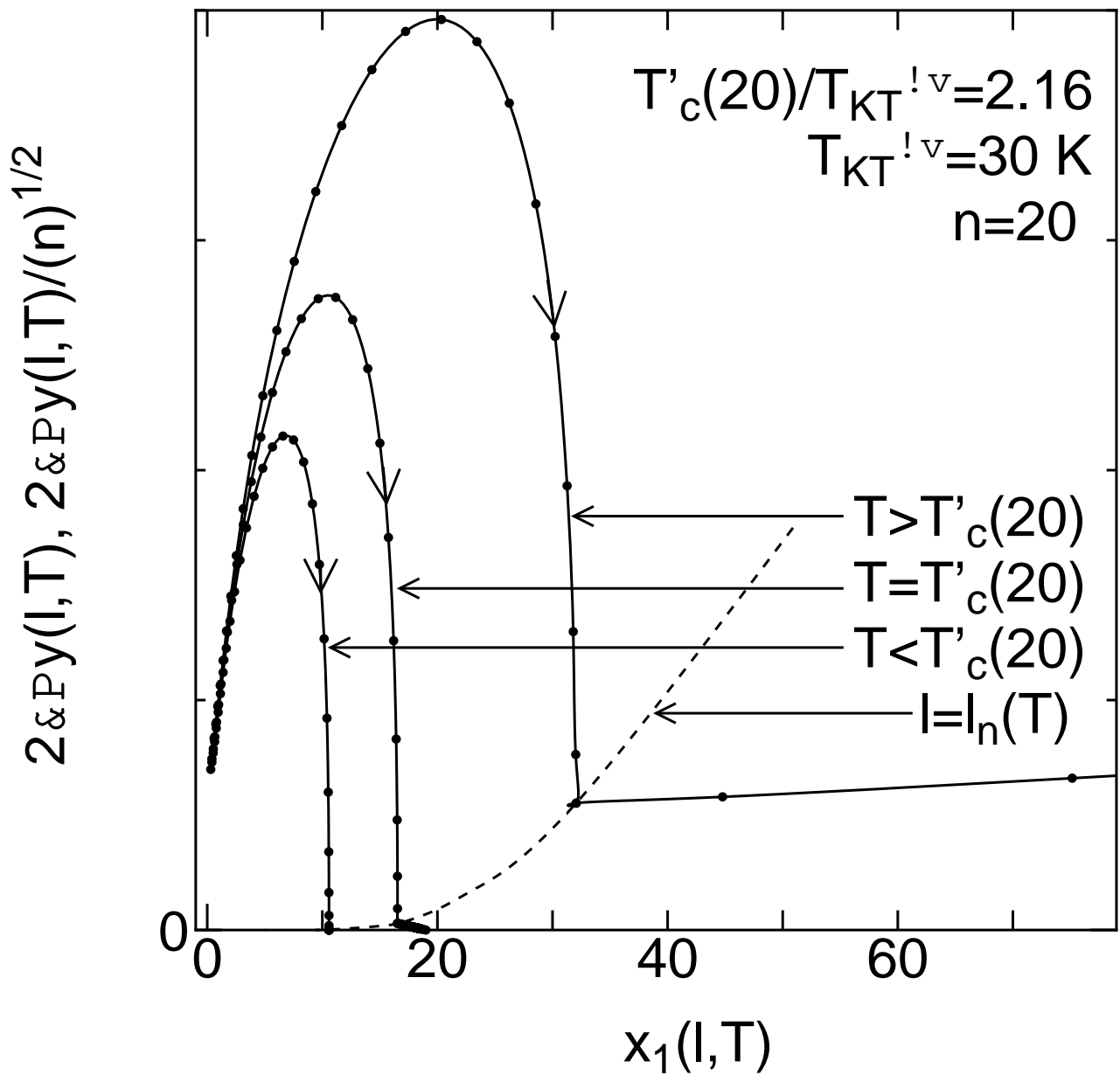
\begin{figure}
\protect%
\caption{%
Renormalization contours of vortex and antivortex pairs. 
For the case (a) the points $ (x(l,T), 2 \pi y(l,T)) $ are shown. 
For the cases (b)-(d), the points $(x_1(l,T),2 \pi y(l,T))$ in the
range $l<l_{n}$ are shown and $(x_1(l,T),2 \pi y(l,T)/
\protect\sqrt{n})$ are in the range $l>l_{n}$. 
The points correspond to intervals of $l$=0.1. The parameter $b$=1.7
is used. For (b)-(d) $K_1/K_{\perp}$=10${}^3$ is used.  
(a) $n$=1, and $K_{\perp}$=0. The curves are with
$T/T_{KT}^{\ast}$=0.5, 1.0, and 1.3 (from left to right).  
The initial points $ (x(0,T), 2 \pi y(0,T)) $ are shown as a dashed curve. 
(b) $n$=3. $T/T_{KT}^{\ast}$=1.40, 1.47, and 1.49. 
(c) $n$=6. $T/T_{KT}^{\ast}$=1.55, 1.58, and 1.62. 
(d) $n$=20. $T/T_{KT}^{\ast}$=2.13, 2.16, and 2.20. 
The transition points from independent vortex pairs
$(x_1(l_n-0,T),2\pi y(l_n-0,T))$ to vortex-string pairs
$(x_1(l_n+0,T),2\pi y(l_n+0,T)/\protect\sqrt{n})$ for $n$=3, 6, and 20
(Fig.~{\protect\ref{fig1}}(b)-(d)) are connected by dashed curves.
 }
\label{fig1}
\end{figure}

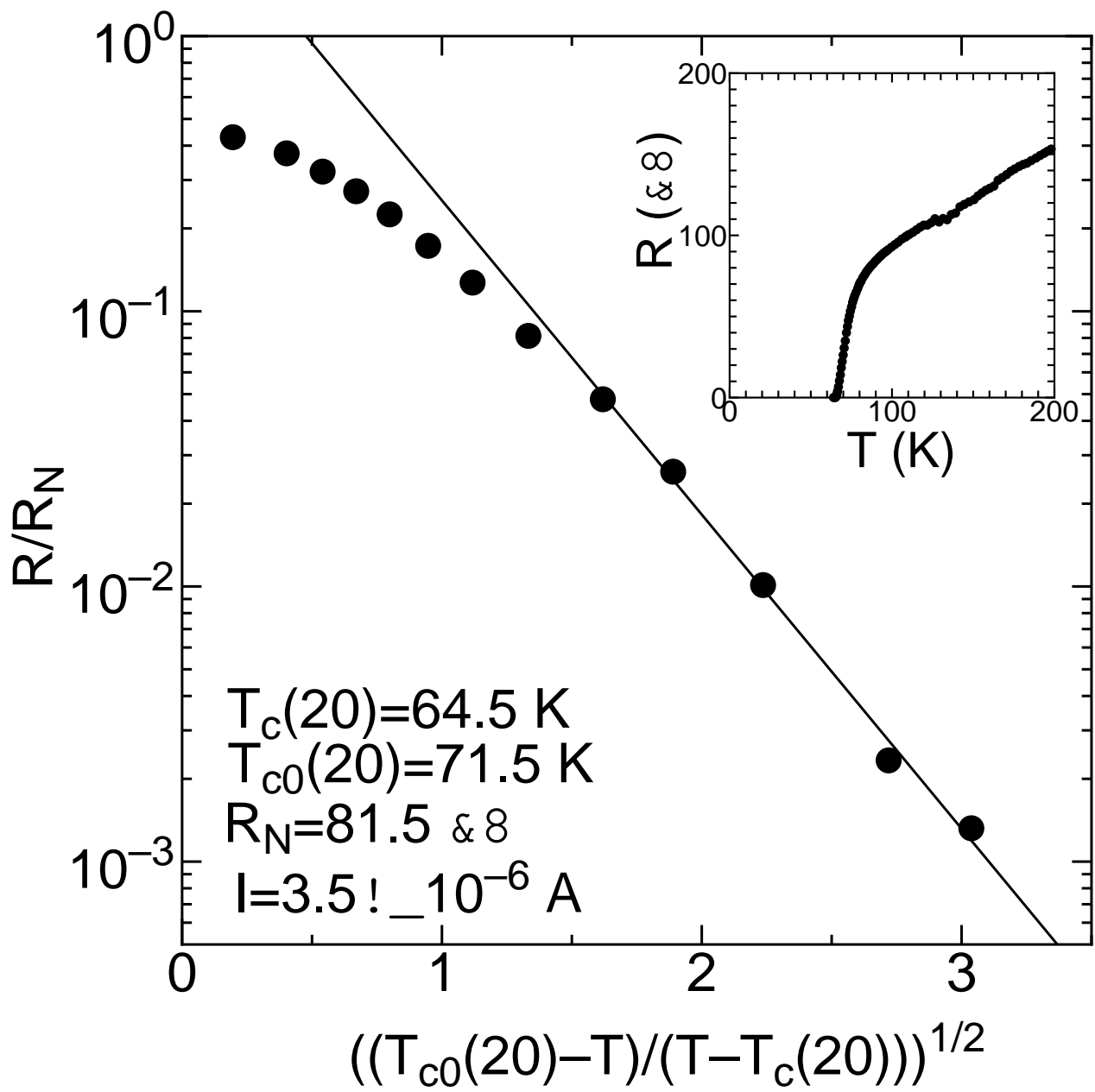
\begin{figure}
\protect%
\caption{ 
Normalized resistance $R/R_N$ as a function of normalized temperature. $T_{c}(20)$=64.5 K and $T_{c0}(20)$=71.5 K are obtained from the inset. 
$R_{N}$ is defined by the normal-state resistance at the onset temperature. 
Inset: Temperature dependence of the resistance for BSCCO film ($n$=20) at $I=3.5 \times 10^{-6}$ A.}
\label{fig2}
\end{figure}

The solutions of the RG equations are shown in Fig.~\ref{fig1}. The RG flow for the case $n$=1 is shown in Fig.~\ref{fig1}(a). 
The parameters $b$ and $T_{c0}(1)$ are extrapolated to the case $n$=1 from those for $n$=3, 6, and 20 shown in Table \ref{table1}. 
The points $(x(0,T),2\pi y(0,T))$ are shown as a dashed curve. 
The points $(x(l,T),2\pi y(l,T))$ follow the flow of 2D XY model as $l$ increases. 
In our numerical solution the RG equation of Eq.~(\ref{xy2}) includes the factor $1/(x(l,T)+1)$ on the right-hand side of Eq.~(\ref{xy2}). 
In the limit $\mid x(l,T) \mid \rightarrow 0$, the solution follows a usual parabolic one.\cite{KT,Halperin} The RG flow starts from $x(0,T'_c(1))$ for $T'_c(1)/T_{KT}^{\ast}$=1.0, and the point converges to $x(l,T'_c(1)) \rightarrow 0$, and $2\pi y(l,T'_c(1)) \rightarrow 0$. For $T>T'_c(1)$, $2\pi y(l,T)$ diverges hyperbolically and then asymptotically linearly to infinity in the limit $x(l,T) \rightarrow \infty$. For $T<T'_c(1)$, $x(l,T)$ is confined within limited range for $2\pi y(l,T)>0$.

RG flows for multilayered thin films are shown in Figs.~1(b)-(d). 
We select the single layer KT-transition temperature as $T_{KT}^{\ast}$=30 K. In these figures, we show the contour of points $(x_1(l,T),2\pi y(l,T))$ for $l<l_n$ and $(x_1(l,T),2\pi y(l,T)/\sqrt{n})$ for $l>l_n$ instead of $(x(l,T),2\pi y(l,T))$ in the case of $n$=1. 
Initially for $l\ll l_n$ the points $(x_1(l,T),2\pi y(l,T))$ are associated with the pairs of independent vortex, and follow the flow of pure 2D XY model. 
As $l$ increases further, the RG flow is deflected downward, {\it i.e.\/}, density of independent vortex is largely reduced, owing to the influence of the $r^2$ interlayer coupling term. 
Then, at $l=l_n$ the flow shows a transition to the region of vortex string. 
The transition occurs continuously or shows a slight cusp depending upon temperature. The points $(x_1(l_n(T)-0,T),2\pi y(l_n(T)-0,T))$ or equivalently $(x_1(l_n(T)+0,T),2\pi y(l_n(T)+0,T)/\sqrt{n})$ for various temperatures are connected as dashed curves in Figs.~1(b)-(d). 
When $l>l_n $, the points $(x_1(l,T),2\pi y(l,T)/\sqrt{n})$ follow the 2D XY RG flow with respect to vortex string. 
Below and above the temperature $T'_c(n)$, the RG flow behaves differently. In the region $T<T'_c(n)$, the flow has limited range, while for $T>T'_c(n)$ the flow diverges parabolically and then asymptotically linearly like 
$2\pi y/\sqrt{n} =x/\sqrt{n}$ to infinity following the 2D KT RG flow with respect to vortex string and the density of dissociated vortex string increases as a function of $l$.
This is the KT transition with respect to vortex string under the influence of interlayer coupling. 
From this, we have the transition temperature $T'_c(n)/T_{KT}^{\ast}$=1.47$\pm$0.1, 1.58$\pm$0.1, and 2.16$\pm$0.1 for the cases $n$=3, 6, and 20, respectively as listed in Table \ref{table1}. 
Note that $x_1(l,T)$ converges to $n-1$ as $y(l,T)\rightarrow 0$ at $T$=$T'_c(n)$. 
The renormalization flow indicates that the vortex string persists below and above $T'_c(n)$.

\section{Sample Preparation and Measurement}

Samples of BSCCO thin films with various thicknesses were prepared by MBE on Nd:YAlO${}_3$ (001) substrates, as described elsewhere.\cite{tsuka,watap,watam} 
All of the films are $c$-axis oriented. 
Current flow within the $ab$ plane was measured so that the circular current, which forms the vortex aligned along the $c$ axis, are the cause of the resistive transition. 
Usually, BSCCO thin films are twinned with respect to the $a$ and $b$ axes. 
On the contrary, our films grown on Nd:YAlO${}_3$ (001) are untwinned.\cite{tsuka,watap,watam} 
We used the untwinned films to reduce the pinning effect of the vortices by the twin boundaries. We have obtained the exact film thickness by the following relation; 1.54 nm $\times$ number of stacked layers. 
The stacked number is estimated from the oscillations due to the Laue function in the x-ray diffraction pattern. 
These samples showed good superconducting transitions.

Electrical measurements were done by D.C. current method with a usual four-terminal method. 
The four contacts ( Au wire with diameter of 25 $\mu $m, Ag paste) were made on the $ab$ plane of the sample. 
During the measurement, care was taken to keep the temperature constant within the error of 0.05 K. 
Errors in the indicated temperatures were kept within 0.1 K by minimizing the difference in measured electrical characteristics obtained during cooling and heating cycles.

\section{Experimental Results and KT transition by Vortex String}

In this section, temperature dependence of resistance $R$ and nonlinear $I-V$ characteristics on thin films at various temperatures near $T_c$ are described. 
These results are interpreted in terms of KT scaling by vortex strings. 
Films with their thicknesses less than $r_0$ should be used in order to eliminate vortex rings. 
To obtain the experimental parameters to resolve the RG equations, we used the data of the film with $n$=20, which is thin enough to exclude the effect of the vortex ring. 
The typical resistance behavior is shown in Fig.~2. 
The sample size along the $ab$ plane is 2.0 mm $\times$ 6.0 mm and the thickness is 30.9 nm. 
The inset of Fig. 2 shows a representative result of the electrical resistance of thin films with $n$=20 as a function of temperature. 
This figure shows the superconducting transition temperature (zero-resistance temperature) $T_c(20)$=64.5 K, the onset temperature $T_{onset}\sim$79 K, and the midpoint temperature $T_{mid}(20)$=71.5 K, which is used as the GL mean-field transition temperature $T_{c0}(20)$.

Applying the pure 2D KT theory \cite{KT,Halperin,Kadin} to vortex string, we have the resistance $R$ that arises from the flow of dissociated free vortex strings at $T>T'_{c}(n)$ as
\begin{equation}
R/R_N \propto \exp\left(-2\sqrt{b{{T_{c0}(n)-T}\over {T-T'_{c}(n)}}}\right).
~(T>T'_{c}(n)) 
\label{RRN}
\end{equation}
Here, $R_N$ is defined by the normal-state resistance at the onset temperature. Note that $T'_c(n)$ is used in place of KT-transition temperature for pure 2D KT theory. 
Equation (\ref{RRN}) is based on the assumption that KT transition occurs owing to the dissociation of vortex-string and antivortex-string pairs as we have shown in Sec. 2. 

Normalized resistance $R/R_N$ is shown on a logarithmic scale as a function of the normalized temperature $\sqrt{{{T_{c0}(20)-T}\over {T-T_c(20)}}}$ in Fig.~2. Data points lie on a straight line around $T_c(20)$. 
The coefficient $b=1.7$ is derived from the gradient of the $R$ vs $T$ curve, using $T_c(20)$=64.5 K, and $T_{c0}(20)$=71.5 K in Eq.~(\ref{RRN}). 
The $n$ dependence of $b$ is rather small as is shown in Table \ref{table1}. 
We extrapolated these values of $b$ for $n>1$ to the value $b$=1.7 for $n$=1 in olving RG equation as was described in Sec.~2. 

According to KT theory, $I-V$ characteristics should exhibit nonlinear power-law behavior $V \propto I^{\alpha (T)}$ owing to the current-induced dissociation of the pairs of vortex string and antivortex string. 
The exponent $\alpha (T)$ of vortex string is given by \cite{KT,Halperin,Kadin}
\begin{equation}
\alpha (T)=1+{{\pi K(l,T)}\over {k_BT}}=1+{{\pi {\hbar}^2 n_s^{2D} } \over {4m^{\ast} \chi(r,T) k_BT}}. \label{at}
\end{equation}
This is based upon our RG analysis where the vortex string persists below and above $T'_c(n)$. According to KT theory, the exponent $\alpha (T)-1$ jumps from 2 to 0 at $T'_{c}(n)$. 
This point determines $T'_c(n)$ as we shall discuss in detail in Sec.~5 (See Eq.~(\ref{tc})).

To see the variation of $\alpha (T)$, the $I-V$ characteristic was measured at various temperatures around $T_c(20)$, using the same sample, as is shown in Fig. 3. 
The $I-V$ curves on a log-log plot yield straight lines over 2 orders of magnitudes in current, indicating power law $V\propto I^{\alpha (T)}$. The gradients of the lines get steeper as the temperature gets lower around $T_c(20)$. 
Figure 4 shows the exponent $\alpha (T)-1$ as a function of temperature obtained from the $I-V$ characteristics  in Fig. 3 and the renormalized coupling constant $(\pi K(l,T))/(k_BT)$ (namely $1/(x(l,T)+1)$) calculated in Sec. 2. 
The filled circles show the measured points. These circles  are compared with the solid curves of the calculated coupling constant. 
Below 66 K, the exponent $\alpha (T)-1$ jumps from 0 to 2 and increases more steeply at lower temperatures. 
At $T\sim T_c(20)$=64.5 K, the exponent $\alpha (T)-1\sim 2$ holds. 

The coupling constant $(\pi K(l,T))/(k_BT)$ of Eq.~(\ref{at}) is drawn as a function of $T$ around $T'_c(20)$ for $l$=3, 4, 5, 10, and 20 in Fig.~4. 
These $l$ regions are, as we have mentioned in Sec.~2 (Table \ref{table1}), in the vortex-string region. 
For a given $l$ and the temperatures higher than $T'_c(20)$, $(\pi K(l,T))/(k_BT)$ drops abruptly. For the larger values of $l$, the gradient gets steeper. 
In the $T$ range below and near $T'_c(20)$, the theoretical curves can be fitted to the experimental results within an experimental error. Above $T'_c(20)$, the calculated coupling constant for the vortex strings of $l$=4 fits the experimental curve best. Then, we can estimate the size of interacting vortex-string pairs as $r={\xi}_{ab}(T)\exp{(3)}\sim$590 nm, near the transition temperature $T'_c(20)$.

The various transition temperatures $T_{c}(n)$, $T'_{c}(n)$, and $T_{c0}(n)$ are shown as functions of $n$ in Fig.~5. To calculate $T'_c(n)$, we have used the experimental values for $T_{c0}(n)$, $b$, ${\lambda}_{ab}(T)$, as are listed in Table \ref{table1}, and $K_1/K_{\perp}=10^3$ (initial). Here $T_{mid}(n)$ is regarded as $T_{c0}(n)$. 
The agreement between the experimental $T_{c}(n)$ and the calculated $T'_{c}(n)$ are satisfactory.

\section{Discussion and Conclusion}

Using the solution of the RG equations, the energies of $U_{vp}(r)$, $U_{ring}(r)$, $U_{st}(r)$ of Eqs.~(\ref{Uvp})-(\ref{Ust}) and the free energies of $F_{vp}(r)$ of Eq.~(\ref{Fvp}) are shown as a function of $r$ for the case $n$=20, at temperatures around $T'_c(20)$ in Fig. 6. 
The dashed curve representing $F_{vp}(r)$ intersects the curve of $U_{st}(r)$ at $r=r_n$. 
We used the renormalized $K_1/K_{\perp}$ obtained in Sec. 2 in Eqs.~(\ref{Uvp})-(\ref{Ust}). 
This figure clearly visualizes that the lowest free-energy excitation changes from that of independent vortex pair to vortex-string pair at $r=r_n$ at all temperatures. 
At large $r$, the free energy of a vortex-string pair is the lowest because of its weakest $r$ dependence. 
At $T=T'_c(20)$, the free energy of a vortex-string pair increases slowly like $\ln r$ because the coupling constant becomes a constant value $K_1\sim 2k_BT'_c(20)/\pi$. 
At $T>T'_c(20)$, it declines because $K_1$ decreases rapidly. 
At $T<T'_c(20)$, although not shown, it behaves similarly as it does at $T=T'_c(20)$. 
For the characteristic length $r_n$ the renormalized $K_1/K_{\perp}$ is used. Notice that the relation $nd<r_n$ still holds after renormalization. 
A cusp of the energy of vortex-string pair at $r=r_n$ appearing in Fig.~6(b) for $T>T'_c(20)$ is due to the large change of $K_1(l,T)$ at the transition from independent vortex pairs to vortex-string pairs that appeared in RG flow in Fig.~1(d). 
The position of $r$=590 nm, at which the $\alpha (T)-1$ vs $T$ relation fits the experimental result best, is located outside the range of Fig.~6. 

Now let us reconsider the initial condition for RG equations of Eqs.~(\ref{xy1}) and (\ref{xy2}) in some detail by taking account of entropy corrections. 
We need some justification for this condition assumed in Sec~2. We are discussing a narrow region near $r\approx {\xi}_{ab}(T)$ in Fig.~6 where energies of the three excitations are close to one another. 
We have a narrow region near $r$=${\xi}_{ab}(T)$, although not recognizable in Fig.~6, where vortex rings are the lowest free-energy excitations but independent vortex pairs are not for the case ${\xi}_{ab}(T)<nd$ and $K_1/K_{\perp}>K_r$. For the case ${\xi}_{ab}(T)>nd$ and $K_1/K_{\perp}>K_p$, we have a narrow region near $r$=${\xi}_{ab}(T)$ where vortex-string pairs are the lowest free-energy excitations but independent vortex pairs are not. 
However, for large anisotropic ratio, the widths of these regions are of the order of 
${\xi}_{ab}(T)(K_{\perp}/K_1)({\xi}_{ab}(T)/d)^2$ or ${\xi}_{ab}(T)\sqrt{(K_{\perp}/K_1)}$ or ${\xi}_{ab}(T)(K_{\perp}/K_1)({\xi}_{ab}(T)/d)$. 
In addition, the energy difference among excitations is also small in these regions. So if we regard that the initial lowest free-energy states are independent vortex for solving RG equations, errors should be small.

Now let us discuss $T'_c(n)$ vs $n$ relation obtained in Sec.~2. 
$T'_{c}(n)$ is defined by the temperature where $x(l,T'_c(n)) \rightarrow 0$ and $y(l,T'_c(n)) \rightarrow 0$ are satisfied as $l \rightarrow \infty $. 
Accordingly, from Eq.~(\ref{x}) and using the relation $K_1(0,T'_c(n))\propto (1-T'_c(n)/T_{c0}(n))$, we have 
\begin{equation}
T'_c(n)={{T_{c0}(n)}\over{1+{{\chi(\infty,T'_{c}(n))}\over n}{{T_{c0}(n)}\over{T_{\Lambda}}}}}, 
\label{tc}
\end{equation}
where the relative dielectric permittivity $\chi(\infty ,T'_c(n))$ is defined as $\chi(\infty ,T'_c(n))=K_1(0,T'_{c}(n))/K_1(\infty ,T'_{c}(n))$. 
$T_{\Lambda}$ is the scaling temperature for $T_{c0}(n)$ defined as 
$T_{\Lambda}=(\pi{\hbar}^2dc^2{\epsilon}_0)/(8k_Be^2{\lambda}^2_{ab}(0))$. 
For the case of $n$=1, Eq.~(\ref{tc}) yields $T'_{c}(1)$=$T^{\ast}_{KT}$. 
Using the parameters for BSCCO \cite{Shiba}, we have $T_{\Lambda}$=223.7 K. 
From Eq.~(\ref{tc}), $T_{c0}(n)$ and $T^{\ast}_{KT}$ set the upper and lower bounds for the $T'_c(n)$ vs $n$ relation, respectively. 
The experimental result of $T_c(n)$ vs $n$ relation shown in Fig.~5 is similar to that of YBCO in a sense that $T_{c0}(n)-T^{\ast}_{KT}$ is of the same order of $T^{\ast}_{KT}$ itself. 
This means that the interplane interaction is equally important as the intraplane KT interaction. 
The difference lies in the shape of $T'_c(n)$ vs $n$ relation. 
This difference arises from the difference in $n$ dependence of $\chi (\infty,T'_{c}(n))$, which in turn originates from the difference in the anisotropy ratio $K_1/K_{\perp}$, as is indicated in Eq.~(\ref{tc}). 
For smaller anisotropy parameters, {\it i.e.\/}, $K_1/K_{\perp} \approx$1, $l_n$ gets smaller, and intervortex interactions are screened largely by vortex string pairs. 
In this case, $T'_c(n)$ approaches that of 2D KT theory with respect to vortex string. 
Matsuda {\it et~al.}'s model corresponds to the case of the limit $l_n$=0 in our model. 
In contrast, for larger anisotropy ratio $K_1/K_{\perp}\gg$1, $l_n$ gets larger and intervortex interactions are screened by independent vortex pairs with $r<r_n$ and by vortex-string pairs with $r>r_n$. 
In this case, $T'_c(n)$ approaches less rapidly that of 2D KT model with respect to vortex string. This is the case for BSCCO and to solve the RG equation is inevitable to obtain $T'_c(n)$ vs $n$ relation.

Now, within the framework of our model, we discuss the absence of dissociation for independent vortex and antivortex pairs. 
In our model, KT transition occurrs in vortex-string region. 
This is due to the absence of $r^2$ term as is indicated in Eq.~(\ref{Ust}) in this region. This can be understood by the following reasoning. 
If we have the interlayer coupling term in the form of $K_{\perp}{(R_c/d)}^{\gamma} (\gamma >0)$, 
the free energy of the system with an isolated independent vortex is given by 
$F_{sv}(R_c)=\pi K_1\ln(R_c/{\xi})+K_{\perp}(R_c/d)^{\gamma}-2k_BT\ln(R_c/{\xi})$. 
The last term is the entropy term and $R_c$ is the sample size along the $ab$ plane. In the limit $R_c \rightarrow \infty$, we have $F_{sv}(R_c)\rightarrow \infty$, irrespective of $T$. This means the absence of vortex dissociation. 
This is also suggested by the RG flow in Fig.~1. The RG curves seem to continue to fall off at large $l$ region if transitions to vortex string were absent. 
Thus, in general, we have bound state for infinite sample as far as interlayer coupling is a more steeply rising function of $r$ than $\ln r$. 
Therefore, the low-lying excitation should necessarily be other than independent vortex pairs in infinite area thin films, {\it i.e.\/}, vortex-string and antivortex-string pairs in thin films.

In the case of finite and very anisotropic samples with limited area along the $ab$ plane, however, KT transition by independent vortex and antivortex pairs may occur.
  Equating the first and the second term of Eq.~(\ref{Uvp}), we have 
$r_b=d \sqrt{(K_1/K_{\perp})\ln (r_b/{\xi}_{ab}(T))}$.\cite{Matsuda93} 
For $r<r_b$, the first $\ln r$ term is larger than the second $r^2$ term. The length $r_b$ becomes large as the anisotropy ratio $K_1/K_{\perp}$ becomes large. If the sample size along the $ab$ plane is less than $r_b$, $r^2$-dependent coupling term is negligible as compared with $\ln r$ term within the $r$ range of the sample size and 2D KT transition can occur. In this case, however, $T_c(n)$ will be independent of $n$ but depend on sample size along the $ab$ plane when this size is near $r_b$. 
We believe that this case is irrelevant for our very thin films of BSCCO, since $T_c(n)$ increases as a function of $n$.

Finally, we will make some comments on the significance of our consequence described so far in bulk high-$T_c$ superconductors. 
Following Matsuda {\it et~al.}, we assign the superconducting transition temperature of bulk BSCCO crystal to 2D/3D crossover temperature $T^{\ast}$. 
This is defined by Hikami and Tsuneto as the temperature, at which $x(r\sim r_b,T^{\ast})$ diverges and given by \cite{HT}
\begin{equation}
{{T^{\ast}-T^{\ast}_{KT}}\over {T^{\ast}_{KT}}}=
\left({{\pi}\over {\ln \sqrt{K_1/K_{\perp}}}}\right)^2. \label{tast}
\end{equation}
In RG flows such as in Fig.~1, the position of $x(r\sim r_b,T)$ where the RG flow starts to fall off goes to infinity at $T=T^{\ast}$. 
In our RG flows of Fig.~1 with temperatures around $T'_c(n)$ these deflecting points are located within finite regions. 
This means that the calculated superconducting transition temperatures $T'_{c}(n)$ are less than $T^{\ast}$. 
These results are consistent with our view that the $T_c$ of our thin film is due to vortex string and that of bulk is 2D/3D crossover. 
If we insert $T^{\ast}\sim$70 K and $T_{KT}^{\ast}$=30 K, we have 
$(T^{\ast}-T^{\ast}_{KT})/T^{\ast}_{KT}\sim$1.3. 
As far as we understand, we should use the renormalized $K_1/K_{\perp}$ in Eq.~(\ref{tast}) as we have obtained $l_n$ by solving RG equations. For example, at $T$=70 K, we have $r_b$=28 nm, $K_1(r_b)/K_{\perp}$=160, and the right-hand side of Eq.~(\ref{tast}) reduces to 1.5. A small discrepancy may be due to the renormalization of $K_{\perp}$ in the bulk.

On experimental side, we have grown very thin films by MBE and verified vortex-string excitations therein. We have used untwinned films exhibiting sharp superconducting transitions as described in Sec. 3. However, more systematic experiment determining the physical parameters of thin films are needed to quantify the model. In particular, superlattices with superconducting layers sandwiched by nonsuperconducting layers are very interesting to study the role of interlayer coupling. The measurement of $T_c(n)$ vs $n$ relation on other high-$T_c$ thin films such as La${}_{2-x}$Sr${}_x$CuO${}_4$ and TBCCO is important for the firm establishment of the concept of vortex string.

So far, we have interpreted our experimental results on superconducting properties of BSCCO thin film in terms of KT transition by vortex string. 
In this view, the influence of pinning centers such as defects can be observed as a perturbation to the KT transition. 
The main reasons that we interpret our experiment by KT transition are following. 
(1) There is a broad transition region in superconducting transition. The transition temperature $T_c$ is far lower than $T_{c0}$. 
(2) Universal jump from 3 to 1 is observed in $\alpha$-$T$ relation. 
(3) The $n$ dependence of $T_c$ is observed in films of various numbers of layers. 
The $T_c$-$n$ relation is the experimental evidence of vortex string. 
In addition, it is believed that the pinning effects are weak in BSCCO superconductors \cite{Los} and that our BSCCO film grown on Nd:YAlO${}_3$ (001) substrates are untwinned as was described in Sec. 3. 
If the transport properties of vortices are not dominated by 
vortex-antivortex interaction but by vortex-defect interaction, these behaviors characteristic of KT transition will not be observed. 
For example, in the presence of high-density defects, the universal jump will be obscured.\cite{onogi}
We showed by our RG scheme that to include $r^2$ dependent interaction necessarily leads to the KT transition by vortex string. 
Comparing the solution of RG equation with the experimental results of $\alpha$-$T$ relation, 
we obtained $l$=4, the logarithm of the size of interacting pairs in the unit of ${\xi}_{ab}(T)$ as was described in Sec. 4. 
In our view, the finite size of independent vortex and antivortex pair or vortex-string and antivortex-string pair represents the distance in which independent vortex and antivortex or vortex-string and antivortex-string can migrate without being influenced by the defects. 
From the $\alpha$-$T$ relation for samples with $n$=3, we have $l\simeq$2.5, shorter than $l$=4 for $n$=20. 
This means that the mean free pass for vortex transport becomes shorter as the film becomes thinner and $T_c(n)$, lower. 
In this case, pinning effects are stronger and vortices are more firmly pinned. This tendency seems physically reasonable, since activation energies for vortices to be freed from pinning are lowered at higher temperatures. 
Note that the value $l\simeq$2.5 is rather close to $l_n$=2.34, which suggests that pinning effect near $T_c(n)$ is fairly strong in the films $n$=3.
In this way within the framework of our view, the effects of the pinning can be considered as a perturbation to the KT transition by vortex string and assessed consistently in terms of finite length of $l$. 
Thus, we believe that our view for our experimental results is basically correct.

In conclusion, we have formulated the KT theory with interlayer coupling to quantify the concept of vortex string. 
We write down the KT RG equations on independent vortex pairs for $r<r_n$ and on vortex-string pairs for $r>r_n$ and then set up the scheme connecting the two regions. KT transition occurred in the vortex-string region with $\ln r$ interstring interaction. 
We believe that the fundamental assumptions in our model are basically correct since $I-V$ characteristics and the $T'_c(n)$ vs $n$ relation in our BSCCO films were interpreted by vortex string. Significance of these results was discussed in brief in relation to bulk high-$T_c$ superconductors.

\acknowledgements
We would like to thank Dr. Onogi, Advanced Research Laboratory, Hitachi, Ltd. for sending us the preprint of Ref.~17 prior to its publication. 
We would also like to thank H. Watanabe for helpful discussions.

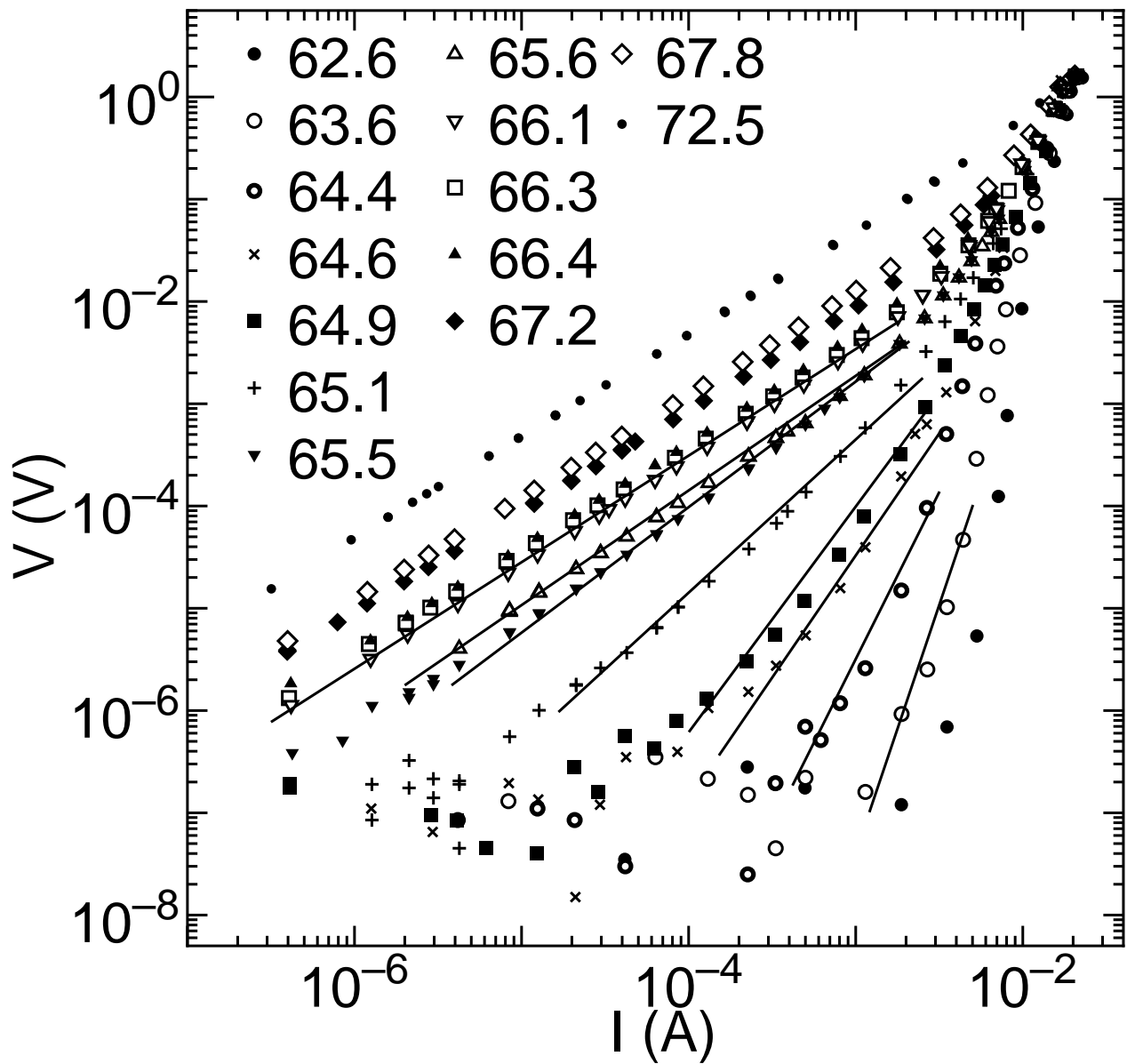
\begin{figure}
\protect%
\caption{ 
Current-voltage characteristics at various temperatures for BSCCO film ($n$=20). Numbers in the figure denote temperatures.}
\label{fig3}
\end{figure}

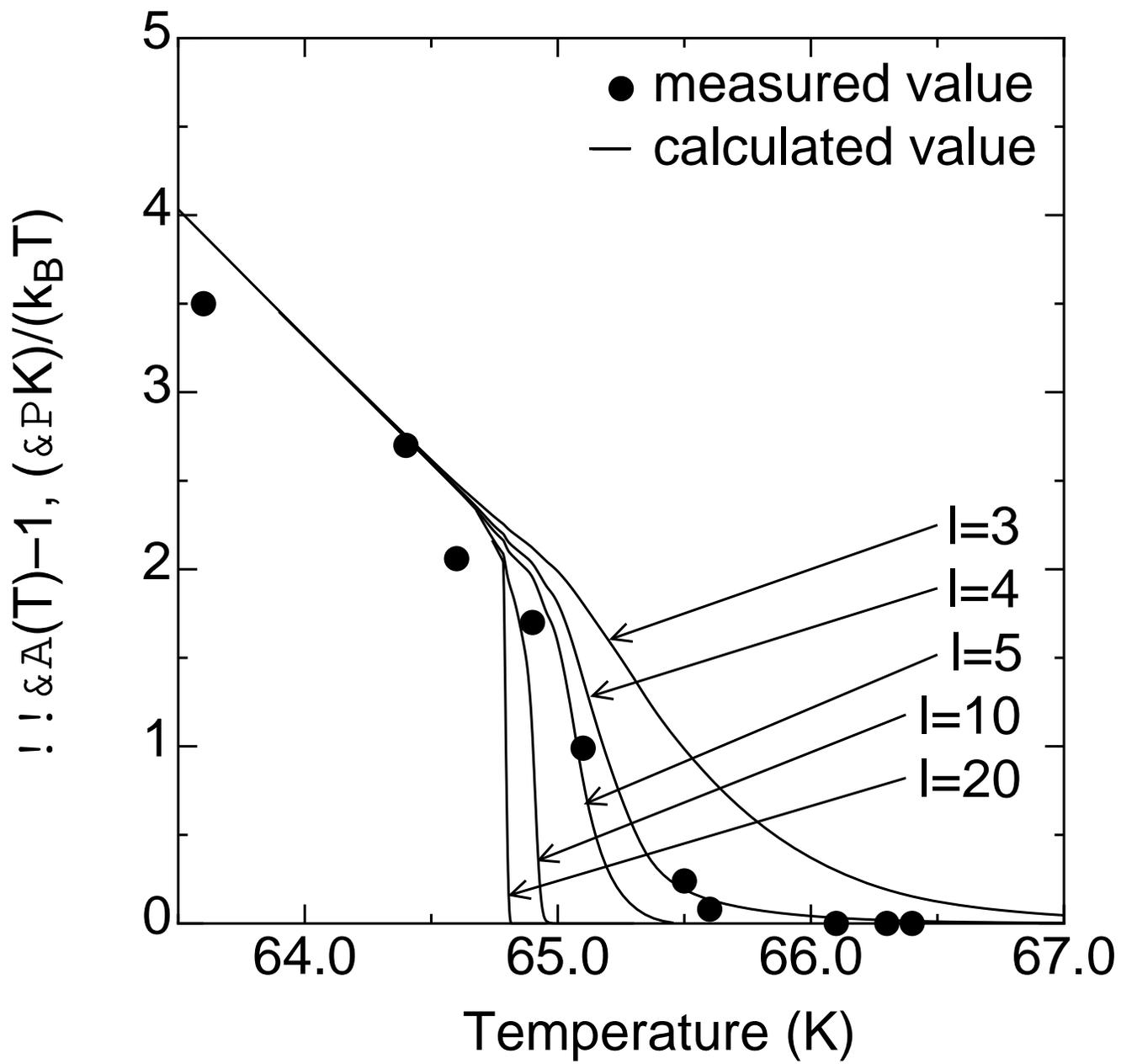
\begin{figure}
\protect%
\caption{ 
The $\alpha (T)-1$ obtained from the $I-V$ characteristics as a
function of temperatures in Fig.~3 and the calculated curves of the
renormalized coupling constant $(\pi K(l,T))/(k_BT)$ (namely
$1/(x(l,T)+1)$) derived from Fig.~{\protect\ref{fig1}}(d). 
The filled circles show the $T$ dependence of measured points of the exponent $\alpha (T)-1$. 
The solid curves are theoretically calculated renormalized
coupling constant.} 
\label{fig4}
\end{figure}

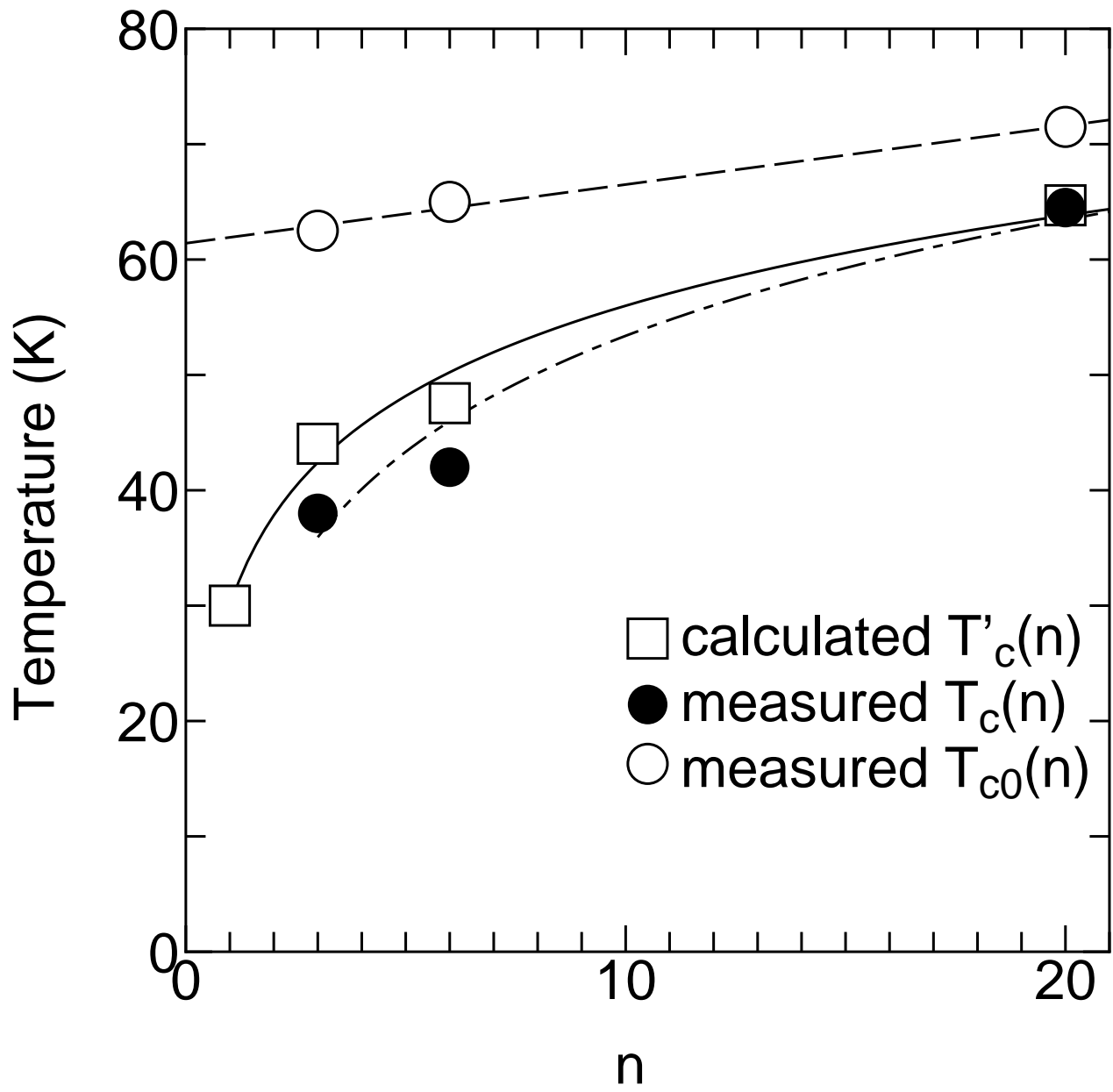
\begin{figure}
\protect%
\caption{ 
Experimental data of $T_{c}(n)$ and $T_{c0}(n)$, and the calculated values $T'_{c}(n)$ as a function of the number of layers $n$. 
Filled circles are the experimental data of superconducting transition temperature $T_{c}(n)$, open circles are the experimental data of GL mean-field transition temperature $T_{c0}(n)$, 
when we assume $T_{c0}(n)=T_{mid}(n)$. 
The open squares are the superconducting transition temperature $T'_{c}(n)$, 
which are obtained from the RG equations (\protect\ref{xy1}) and (\protect\ref{xy2}). 
Data points of  $T_{c}(n)$ and $T_{c0}(n)$ for $n$=3 and 6 half unit-cell thin films are taken from Refs.~20 and 21. 
Solid, dot-dashed, and dashed curves are guides to the eyes for $T'_{c}(n)$, $T_{c}(n)$, and $T_{c0}(n)$, respectively.}
\label{fig5}
\end{figure}

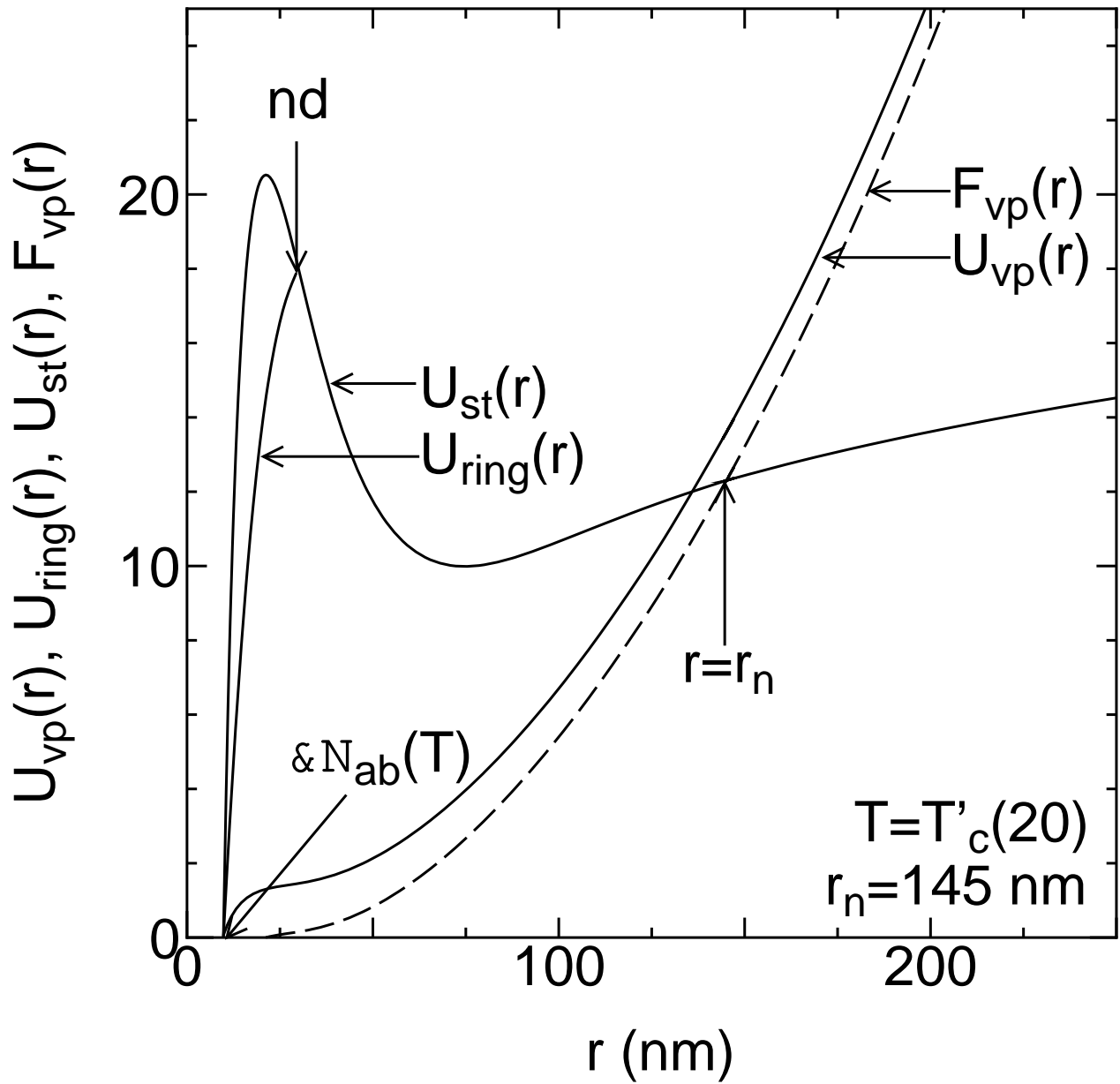
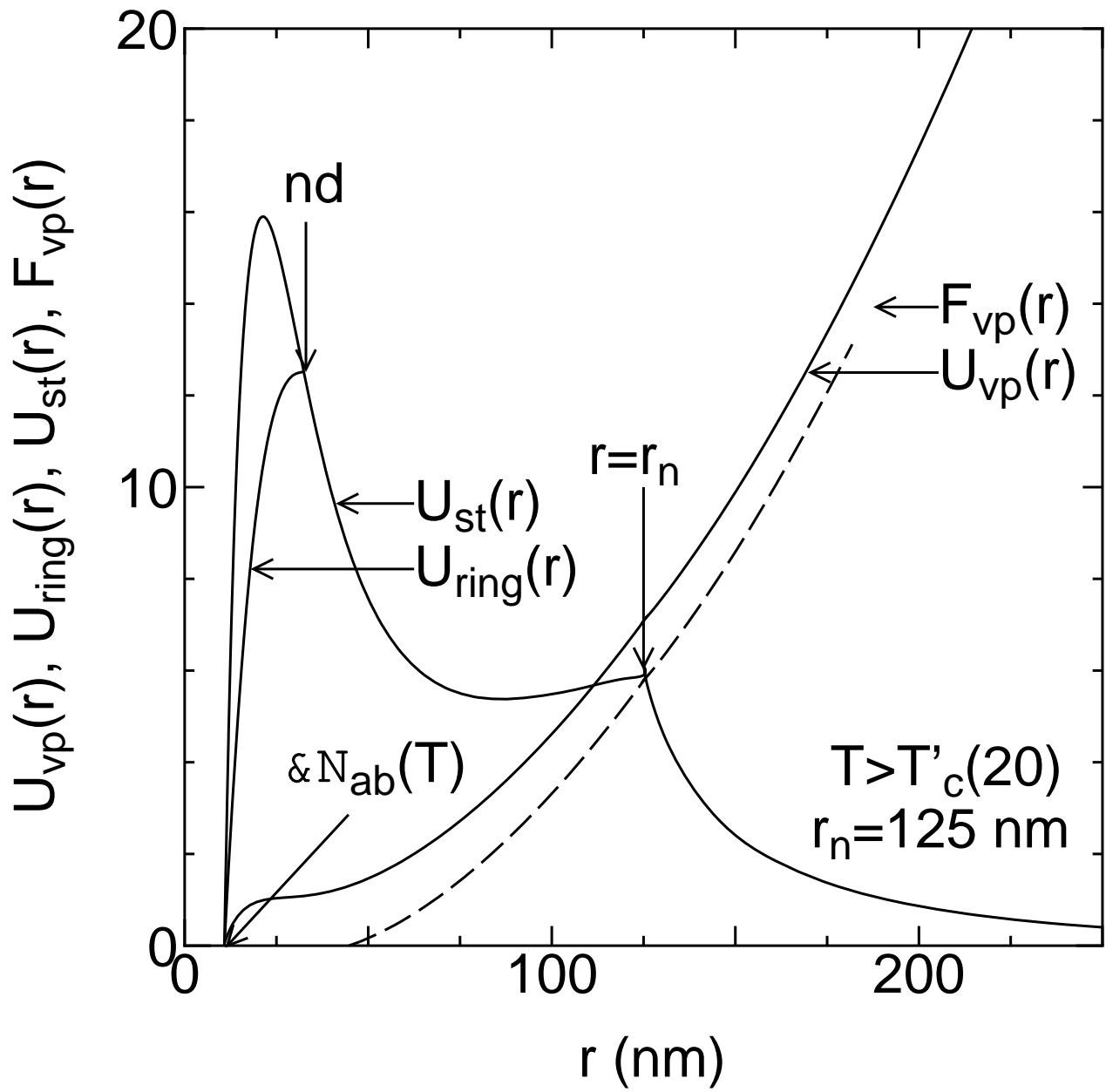
\begin{figure}
\protect%
\caption{
The energies of three types of excitations, 
$U_{vp}(r)$, $U_{ring}(r)$, and $U_{st}(r)$ and the free energy $F_{vp}(r)$ as a function of $r$ in the unit of $k_BT$ for $n$=20. 
(a) $T=T'_c(20)$. (b) $T>T'_c(20)$ ($T/T_{KT}^{\ast}$=2.20). 
$U(r)$ are shown as solid curves and $F_{vp}(r)$, as dashed curves. 
The renormalized $K_1(r,T)$ is used. Their intersections at $nd$ and $r_{n}$ are shown.}
\label{fig6}
\end{figure}

\begin{table}
\protect%
\caption{
List of parameters for BSCCO thin films. ${\xi}_{ab}(T_c(n))$, $l_{n}(T'_c(n))$, $b$, $T_{c0}(n)$, $T_{mid}(n)$, $T_{c}(n)$, $T'_{c}(n)$, and $T'_{c}(n)/T^{\ast}_{KT}$ for $n$=1, 3, 6, and 20. 
${\xi}_{ab}(T_{c}(n))$ are calculated using ${\xi}_{ab}(0)$=3 nm and experimental values of $T_{c}(n)$ and $T_{c0}(n)$. 
$l_n(T'_{c}(n))$ are calculated using Eq.~(\protect\ref{rrnn}) and renormalized
$K_1(l_n(T),T)$, and $d$=1.54 nm. The initial value of
$K_1(l=0,T)/K_{\perp}(l=0,T)$=10${}^3$ is calculated using
${\xi}_{c}(0)$=0.1 nm.\protect\cite{Naughton}  
$b$ for $n$=20 is taken from $R$ vs $T$ relations in Fig.~2. 
$T_{c0}(n)$ or $T_{mid}(n)$ for $n$=20 is taken from the inset of
Fig.~{\protect\ref{fig2}}.
$T_{c}(n)$ for $n$=20 is determined from $R$ vs $T$ relations in the
inset of Fig.~{\protect\ref{fig2}}.
$T'_{c}(n)$ is calculated from RG equations (\protect\ref{xy1}) and
(\protect\ref{xy2}) using experimental values listed in this Table.
$T^{\ast}_{KT}$ is taken as $T^{\ast}_{KT}$=30 K. The parameters
$T_{c}(n)$, $T_{c0}(n)$, and $b$ for $n$=3 and 6 are taken from
Refs.~20 and 21.  
$T_{c0}(n)$ and $b$ for $n$=1 are extrapolated from those for $n$=3,
6, and 20. For the definition of parameters, see the text.} 
\begin{tabular}{lllll}
$n$                          & 1 & 3 & 6 & 20 \\
\tableline
 ${\xi}_{ab}(T_c(n))$\ (nm)  & 4.2 & 4.8 & 5.2 & 9.6 \\
 $l_n(T'_c(n))$              &  (0) & 2.34 & 2.54 & 2.70 \\
 $b$                         & (1.7) & 1.4 & 1.7 & 1.7 \\
 $T_{c0}(n), T_{mid}(n)$\ (K)& (62) &  62 & 65 & 71.5 \\
 $T_c(n)$\ (K)               &  &  38 & 43 & 64.5 \\
 $T'_c(n)$\ (K)              &  (30) & 44.1$\pm$3 & 47.4$\pm$3 & 64.8$\pm$3 \\
 $T'_c(n)/T^{\ast}_{KT}$   &(1) & 1.47$\pm$0.1 & 1.58$\pm$0.1 & 2.16$\pm$0.1 \\
\end{tabular}
\label{table1}
\end{table}

\end{document}